\pdfoutput=1
\documentclass[iop,twocolumn,apjl,letterpaper]{emulateapj}

\usepackage{lipsum}
\usepackage{graphicx}
\usepackage{amsmath}
\usepackage{rotating}
\newcommand{\pomega}{\varpi}

\begin{document}
\title{A Primer on Unifying Debris Disk Morphologies}

\author{Eve J. Lee\altaffilmark{1}, Eugene Chiang\altaffilmark{1,2}}
\altaffiltext{1}{Department of Astronomy, University of California Berkeley, Berkeley, CA 94720-3411, USA; evelee@berkeley.edu, echiang@astro.berkeley.edu}
\altaffiltext{2}{Department of Earth and Planetary Science, University of California Berkeley, Berkeley, CA 94720-4767, USA}

\begin{abstract}
A ``minimum model'' for debris disks consists of a narrow ring
of parent bodies, secularly forced by a single planet on a possibly
eccentric orbit, colliding to produce dust grains that are perturbed
by stellar radiation pressure. We demonstrate how this minimum model
can reproduce a wide variety of disk morphologies imaged in scattered
starlight. Five broad categories of disk shape can be captured: ``rings,''
``needles,'' ``ships-and-wakes,'' ``bars,'' and ``moths (a.k.a.~fans),''
depending on the viewing geometry. Moths can also sport
``double wings.'' We explain the origin of morphological features
from first principles, exploring the dependence on planet eccentricity,
disk inclination dispersion, and the parent body orbital phases at
which dust grains are born. A key determinant in disk appearance
is the degree to which dust grain orbits are apsidally aligned.
Our study of a simple steady-state (secularly relaxed)
disk should serve as a reference for more detailed models
tailored to individual systems.
We use the intuition gained from our guidebook of disk morphologies
to interpret, informally, the images of a number of real-world debris disks. 
These interpretations suggest that the farthest reaches of planetary systems 
are perturbed by eccentric planets, possibly just a few Earth masses each.
\end{abstract}

\section{Introduction}
\label{sec1}

Orbiting stars $\gtrsim 10$ Myr old,
``debris disks'' are
thought to trace the aftermath of planet formation
(for a review, see \citealt{wyatt08}). 
By definition, they are composed of optically thin
dust grains, generated from the collisional
attrition of larger parents. 
Observations of debris disks shed light
on the size distribution and velocity dispersion
of constituent bodies (e.g., \citealt{shannon11};
\citealt{pan12}; and references therein),
and by extension the processes by which
planetesimals and planetoids build up and grind down.
Debris disks may also serve as signposts for embedded planets
(e.g., \citealt{mouillet97}; \citealt{rodigas14}; 
\citealt{nesvold16}; and references therein).

The morphologies of debris disks are coming
into increasingly sharp resolution with the advent
of extreme adaptive optics instruments, including 
the {\it Gemini Planet Imager} \citep[{\it GPI};][]{macintosh14},
{\it SPHERE} \citep{beuzit08}, and {\it SCExAO} \citep{jovanovic15}.
In past and present observing campaigns,
a variety of disk shapes have been uncovered,
some featuring warps \citep[e.g.,][]{heap00,apai15,mmb15,wang15} 
and eccentric rings \citep[e.g.,][]{kalas05fom,wahhaj14,perrin15},
and others evoking ``moths'' 
\citep[e.g.,][]{hines07,maness09,ricarte13}
and ``needles'' \citep[e.g.,][]{kalas07,kalas15}.
Some imaged features have even been observed to 
vary with time \citep{boccaletti15}.
\citet{schneider14} present a beautiful
compilation of debris disk images taken with
the {\it Hubble Space Telescope} ({\it HST}).

Disk structures that are non-axisymmetric are especially
intriguing because they hint at gravitational sculpting
by planets (assuming disk self-gravity is negligible;
see, e.g., \citealt{jalali12} for a contrarian view).
Foundational work was done by \citet{wyatt99},
who calculated how one or more planets on eccentric,
inclined orbits imprint ellipticities and warps onto debris disks.
The planetary perturbations treated by these authors 
are secular, i.e., orbit-averaged in the sense that the gravitational
potential
presented by each planet is that of a smooth, massive wire
(see also the textbook by \citealt{murray_dermott}).
Mean-motion commensurabilities with a planet can also shape disks 
by truncating them in a chaotic zone of overlapping
first-order resonances
\citep[e.g.,][]{wisdom80,quillen06_fom,pearce14,nesvold15_gap}. 
Individual resonances can also,
in principle, trap disk particles
and clump them azimuthally \citep[e.g.,][]{kuchner03,stark09}. 
Such resonant clumps, moving at pattern speeds that typically differ
from local Kepler frequencies, have yet to be confirmed
in extrasolar debris disks. 
The preponderance of evidence shows that
debris disks are smooth \citep[e.g.,][]{hughes12},
suggesting that secular effects dominate their 
appearance.

We offer here a systematic exploration of the morphologies
of planet-perturbed debris disks, as imaged in scattered
starlight. We focus on what is arguably the simplest possible
scenario: a narrow ring of parent bodies forced
secularly by a single planet, producing dust grains that
are propelled outward by stellar radiation pressure.
Our work builds on \citet{wyatt99} by
supplying synthetic scattered light images of disks
viewed from all possible directions. For all its simplicity,
the model contains a surprisingly large variety of
morphologies, and we will assess, in a qualitative way, the extent
to which the observed real-world diversity of shapes
(rings, flares, moths, needles, and the like)
may be attributed to differences in viewing geometry;
in other words, we explore a ``unification'' model for debris disks,
by analogy with unification models for active galactic
nuclei. We do not expect our model to be able
to explain every detail of resolved disk images,
but submit our work as a starting point for interpreting
those images: a baseline reference that can guide more
sophisticated theories.

Our paper is straightforward.
After describing the model elements and 
computational procedure (Section \ref{sec2}),
we present synthetic scattered light images
(Section \ref{sec3})
and compare them informally to actual systems
(Section \ref{sec4}).
Our aim is to provide a primer on debris disk
morphology: to explain features from first principles,
and develop intuition for mapping scattered light
observations to the underlying parent 
disks and attendant planets.
This paper is intended as a more general
expansion of ideas discussed by 
Esposito et al.~(submitted)
to explain the moth-like
morphology presented by HD 61005
(see also \citealt{fitz11} for the original proposal).

\section{Model}
\label{sec2}

We posit a planet of mass 
$M_{\rm planet} = 10 M_\oplus$ 
on an orbit with semi-major axis $a_{\rm planet} = 30$ AU
and eccentricity $e_{\rm planet} \in (0, 0.25, 0.7)$
about a star with mass $M_\ast = 1 M_\odot$. 
The planet's orbit lies in the reference ($x$-$y$) plane,
with its longitude of periapse $\pomega_{\rm planet} = 0$
(the planet's periapse is located on the $x$-axis).

Debris disk bodies are of two kinds:
parent bodies and dust particles. The latter
are spawned from the former. Parent bodies (subscripted p)
are located exterior to the planet's orbit and number 
$N_{\rm p} = 1000$ in all.
They have semi-major axes distributed uniformly
from just outside the planet's chaotic zone
\citep{wisdom80,quillen06_fom,quillen06,chiang09,nesvold15_gap},
\begin{equation}
    a_{\rm p,inner} = a_{\rm planet} \left[1 + 2 (M_{\rm planet}/M_\ast)^{2/7}\right] \,,
\label{eq:ain}
\end{equation}
to a value 10\% larger,
\begin{equation}
  a_{\rm p,outer} = 1.1 a_{\rm p,inner} \,.
\end{equation}
Thus our debris disks are really debris
rings, as inspired by the narrow belts observed in,
e.g., HR 4796A, Fomalhaut, AU Mic, and the Kuiper belt.
For the highest value of $e_{\rm planet} = 0.7$ that we
consider, equation (9) of \citet{pearce14} 
is more accurate and gives a value for
$a_{\rm p,inner} - a_{\rm planet}$
that is $\sim$2 times larger
than the one predicted by our equation (\ref{eq:ain});
we neglect this correction for simplicity.

A parent body's eccentricity vector --- a.k.a.~its
Runge-Lenz vector, which points toward periapse and has a
length equal to the eccentricity --- 
is the vector sum of its
forced and free eccentricities \citep[e.g.,][]{murray_dermott}. 
The forced eccentricity vector is computed
from Laplace-Lagrange (L-L) secular theory; in the
one-planet case which we are considering, the forced
vector points parallel to the planet's 
eccentricity vector (i.e.,
in the positive $x$-direction), and has a magnitude 
specific to the body's orbital semi-major axis:
\begin{equation}
e_{\rm p,forced} = \frac{b_{3/2}^{(2)} (a_{\rm planet}/a_{\rm p})}{b_{3/2}^{(1)} (a_{\rm planet}/a_{\rm p})} \,e_{\rm planet} \,,
\end{equation}
where the $b$'s are the usual Laplace coefficients. 
As $a_{\rm planet}/a_{\rm p} \rightarrow 1$, $e_{\rm p,forced} \rightarrow e_{\rm planet}$.
The components of the free eccentricity vectors,
as resolved in
$(h,k) \equiv (e \sin \pomega, e\cos \pomega)$ space,
are
\begin{align}
h_{\rm p,free} &= e_{\rm p,free} \sin \pomega_{\rm p,free} \\
k_{\rm p,free} &= e_{\rm p,free} \cos \pomega_{\rm p,free}
\end{align}
where $\pomega_{\rm p,free}$ is a uniform deviate
between 0 and $2\pi$ rad, and
$e_{\rm p,free}$ is a uniform
deviate that extends from 0 to 0.02.
The value of $e_{\rm p,free}$ measures
the random velocity dispersion,
which in turn depends on how bodies collide
and are gravitationally stirred 
(processes not modeled here;
see, e.g., \citealt{pan12}).
Total parent body eccentricities
are such that no parent body crosses the planet's orbit;
see \citet{chiang09} for
numerical $N$-body integrations verifying
orbital stability for parameters similar to those used here.
That $\pomega_{\rm p,free}$ ranges
uniformly from 0 to $2\pi$ assumes that parent bodies
are secularly relaxed; for our chosen parameters
($M_{\rm planet}$, $a_{\rm planet}$, $a_{\rm p,inner}$,
$a_{\rm p,outer}$), differential precession
timescales across the parent ring are of order a 
couple of Myrs,
shorter than typical debris disk ages of tens of Myrs.
To summarize, the parent bodies occupy, in the mean,
a narrow elliptical
ring located just outside the planet's elliptical orbit
and apsidally aligned with it.\footnote{The low-order 
Laplace-Lagrange (L-L) secular theory which we use is quantitatively
inaccurate at high $e_{\rm planet}$ but should be
qualitatively correct. \citet{pearce14} find 
good correspondence between their $N$-body integrations
and L-L theory for $e_{\rm planet}$ as high as $\sim$0.8,
provided the planet's orbit lies
within $\sim$20$^\circ$ of the 
parent disk, as it does for all our models.} 

Parent body inclination vectors,
resolved in $(p,q) \equiv (i \sin \Omega, i \cos\Omega)$
space, where $i$ is inclination and $\Omega$ is
the longitude of ascending node, behave analogously
to eccentricity vectors.
For our one-planet case, the forced
inclination vector is the zero vector: forced orbits
are co-planar with the planet's orbit. Therefore parent body
inclination vectors equal their free values:
\begin{align}
p_{\rm p,free} &= i_{\rm p,free} \sin \Omega_{\rm p,free} \\
q_{\rm p,free} &= i_{\rm p,free} \cos \Omega_{\rm p,free}
\end{align}
where $\Omega_{\rm p,free}$
is a uniform deviate between 0 and $2\pi$ rad
(this assumes the system is secularly relaxed; see above),
and $i_{\rm p,free}$
is a uniform deviate between 0 and 0.02 rad.

\begin{figure*}[!ht]
    \centering
    \includegraphics[width=\textwidth]{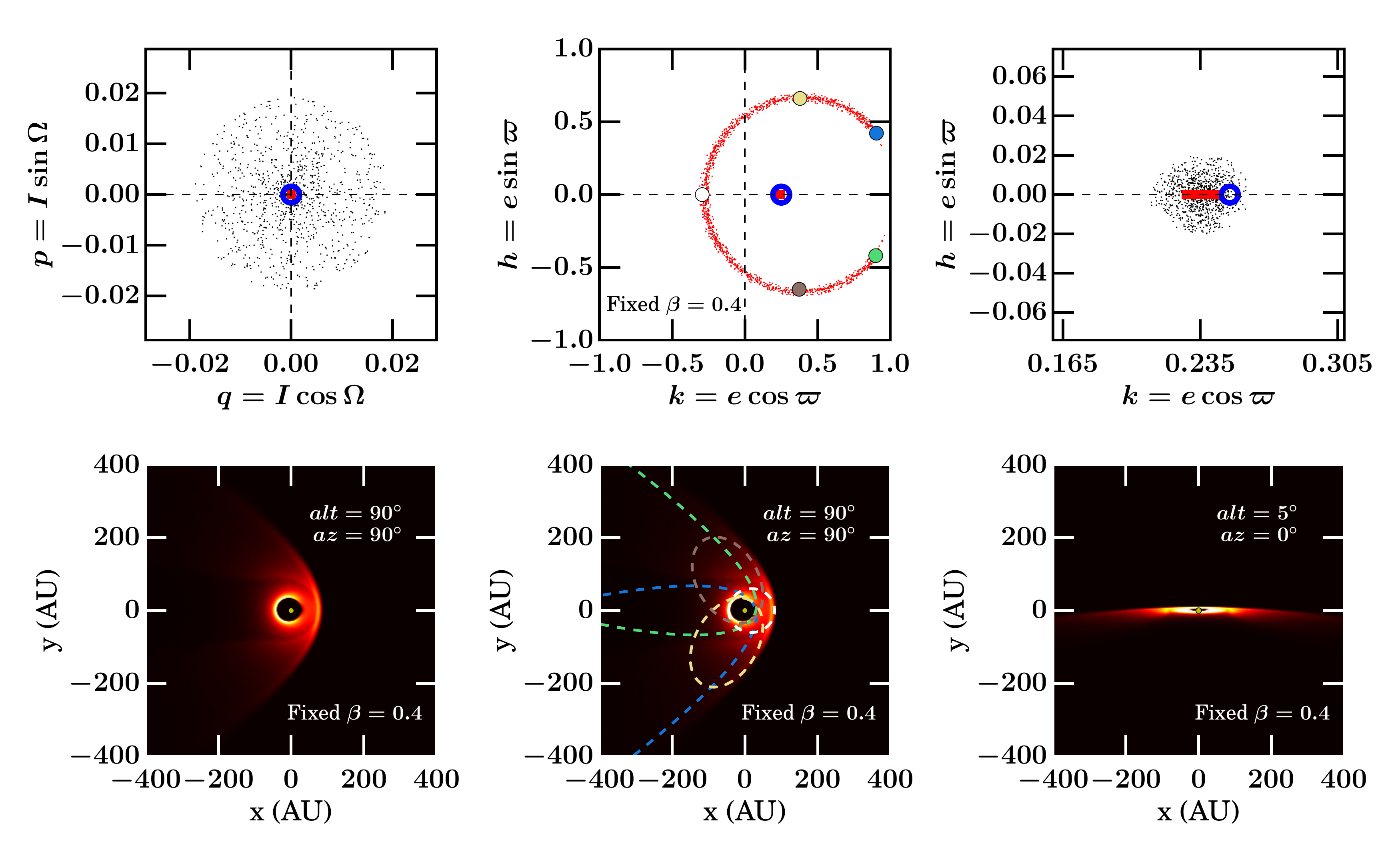}
    \caption{Elements of a sample model for which $e_{\rm planet} = 0.25$,
$\max e_{\rm p, free} = \max i_{\rm p,free} = 0.02$ (in radians),
and --- for this figure alone --- a fixed $\beta = 0.4$ for dust particles.
{\it Top row}: Inclination and eccentricity vector components of the planet
(blue open circle), parent bodies (black points), and dust particles
(red points). Forced eccentricities of parent bodies are shown as a red bar;
full eccentricities differ from their forced values
by up to $\max e_{\rm p,free}$ (top right panel). Similarly, full
inclinations differ from their forced values by up to $\max i_{\rm p,free}$ (the half thickness of the disk).
Because stellar radiation pressure does not alter orbital
inclination, dust particle and parent body inclinations are identical
(black points overlie red points in the top left panel). {\it Bottom row}: 
Synthetic scattered light images for this disk seen face-on ($alt = 90$ deg,
bottom left and middle panels) and seen 5 deg above the planet's orbital
plane ($alt = 5$ deg, $az = 0$, bottom right panel).
The scattered light features in the
face-on (a.k.a.~polar) view can be understood from an underlying
``skeleton'' of representative dust grain orbits,
shown in matching colors
in top and bottom middle panels. The nearly edge-on
view in the right panel is such that the planet's apoapse
points toward the observer.}
    \label{fig1}
\end{figure*}

To launch dust particles (subscripted d)
from parent bodies, 
we first randomly draw, for a given parent body's orbit,
$N_{\rm launch} = 100$
true anomalies. These true anomalies mark the launch
positions for dust grains; for simplicity, we draw the
$N_{\rm launch}$ true
anomalies from a uniform distribution (on the one hand,
periastron may be favored for collisions because
orbital velocities are higher there, but on the other
hand, apastron may be favored because parent bodies spend
more time there; we discuss the effects of different 
choices for the distribution of launch sites in 
Section \ref{sec3}).
At every true anomaly,
a dust particle orbit is created whose instantaneous
velocity at that position matches the parent body's
instantaneous velocity, and whose radiation
$\beta$ --- the ratio
of the force of stellar radiation pressure
to that of stellar gravity --- is drawn randomly from a
distribution to be given below.
To quantify our statements so far, the orbital elements of each
dust grain orbit are given by:
\begin{align}
\label{eq:a_d}
a_{\rm d} &= \frac{a_{\rm p} (1-e_{\rm p}^2) (1-\beta)}{1-e_{\rm p}^2-2\beta(1+e_{\rm p}\cos f_{\rm p})} \\
\label{eq:e_d}
e_{\rm d} &= \frac{\sqrt{e_{\rm p}^2+2\beta e_{\rm p} \cos f_{\rm p} + \beta^2}}{1-\beta} \\
\label{eq:om_d}
\omega_{\rm d} &= \omega_{\rm p} + \arctan\left(\frac{\beta\sin f_{\rm p}}{e_{\rm p} + \beta\cos f_{\rm p}}\right) \\
\label{eq:i_d}
i_{\rm d} &= i_{\rm p} \\
\label{eq:Om_d}
\Omega_{\rm d} &= \Omega_{\rm p}
\end{align}
where $\omega$ is the argument of periapse and $f_{\rm p}$ is the 
parent body's true anomaly at launch.

The $\beta$-distribution is related to the assumed size
distribution of dust grains. If the latter derives
from a standard collisional cascade and obeys, e.g.,
a Dohnanyi distribution $dN/ds \propto s^{-7/2}$
for particle size $s$,
then $dN/d\beta \propto \beta^{3/2}$,
under the assumption that dust particles present
geometric cross sections to radiation pressure ($\beta \propto 1/s$).
But a conventional cascade 
underestimates the number of grains whose sizes
are just shy of the radiation blow-out size.
These grains are on especially
high-eccentricity and high-semi-major-axis orbits,
avoiding interparticle collisions
by spending much of their time
away from the dense parent body ring. 
Their actual lifetimes against
collisional destruction, and by extension their steady-state
population, are underestimated by a standard
cascade. 
We correct for this effect by scaling the number of dust
grains in a given size bin to their orbital period $P_{\rm d}$,
which is longer at higher $\beta$.
This same scaling is used by \citet[][see their Figure 3]{strubbe06}
who show that it correctly reproduces the surface brightness profiles
of collision-dominated debris disks like AU Mic.
Our $\beta$-distribution therefore scales as
\begin{align}
dN/d\beta & \propto \beta^{3/2} \times P_{\rm d} \nonumber \\
& \propto \beta^{3/2} \frac{(1-\beta)^{3/2}}{ \left[ 1 - e_{\rm p}^2 - 2\beta (1+ e_{\rm p} \cos f_{\rm p}) \right]^{3/2}} \label{eq:orbcorr}
\end{align}
where we have used $P_{\rm d} \propto a_{\rm d}^{3/2}$
and equation (\ref{eq:a_d}).
The $\beta$-distribution extends 
from $\beta_{\rm min} = 0.001$ to a maximum value
$\beta_{\rm max}$ corresponding to a marginally
bound (zero energy; $e_{\rm d}=1$) orbit.
Each value of $\beta_{\rm max}$ is specific
to a given launch position and velocity.
The $\beta$-distribution given by (\ref{eq:orbcorr})
is very top-heavy;
most grains have $\beta$ near
the maximum value
\begin{equation}
\beta_{\rm max} = \frac{1-e_{\rm p}^2}{2(1+e_{\rm p} \cos f_{\rm p})}\, .
\label{eq:betamax}
\end{equation}

Along each dust particle orbit, 
we lay down, at random,
$N_{\rm dust-per-orbit} = 100$ dust particles.
Their mean anomalies are uniformly
distributed but their true anomalies are not; 
dust particles concentrate near apoapse, following
Kepler's equation. 
The dust particles,
numbering $N_{\rm d} = N_{\rm p} \times N_{\rm launch} 
\times N_{\rm dust-per-orbit} = 10^7$ in all,
are projected onto the sky plane of a distant
observer and used to synthesize a scattered light image.
The sky plane of 800 $\times$ 800 AU, centered on the star,
is divided into 800 $\times$ 800 square cells, and each dust particle
contributes, to the cell in which it is located,
a surface brightness
proportional to $\phi(g,\theta)/(\beta^2 r^2)$. 
Here $1/\beta^2$ accounts for the scattering cross
section for each grain (assumed geometric),
$r$ is the distance between the dust particle
and the host star, and $\phi(g,\theta)$ is the 
Henyey-Greenstein scattering
phase function for asymmetry parameter $g = 0.5$
and $\theta$ equal to 
the angle between the dust particle and the observer
with the vertex at the star.
Multiple scattering of photons is neglected; this
is a safe assumption insofar as debris disks
are optically thin.

\begin{figure*}[!ht]
    \includegraphics[width=\textwidth]{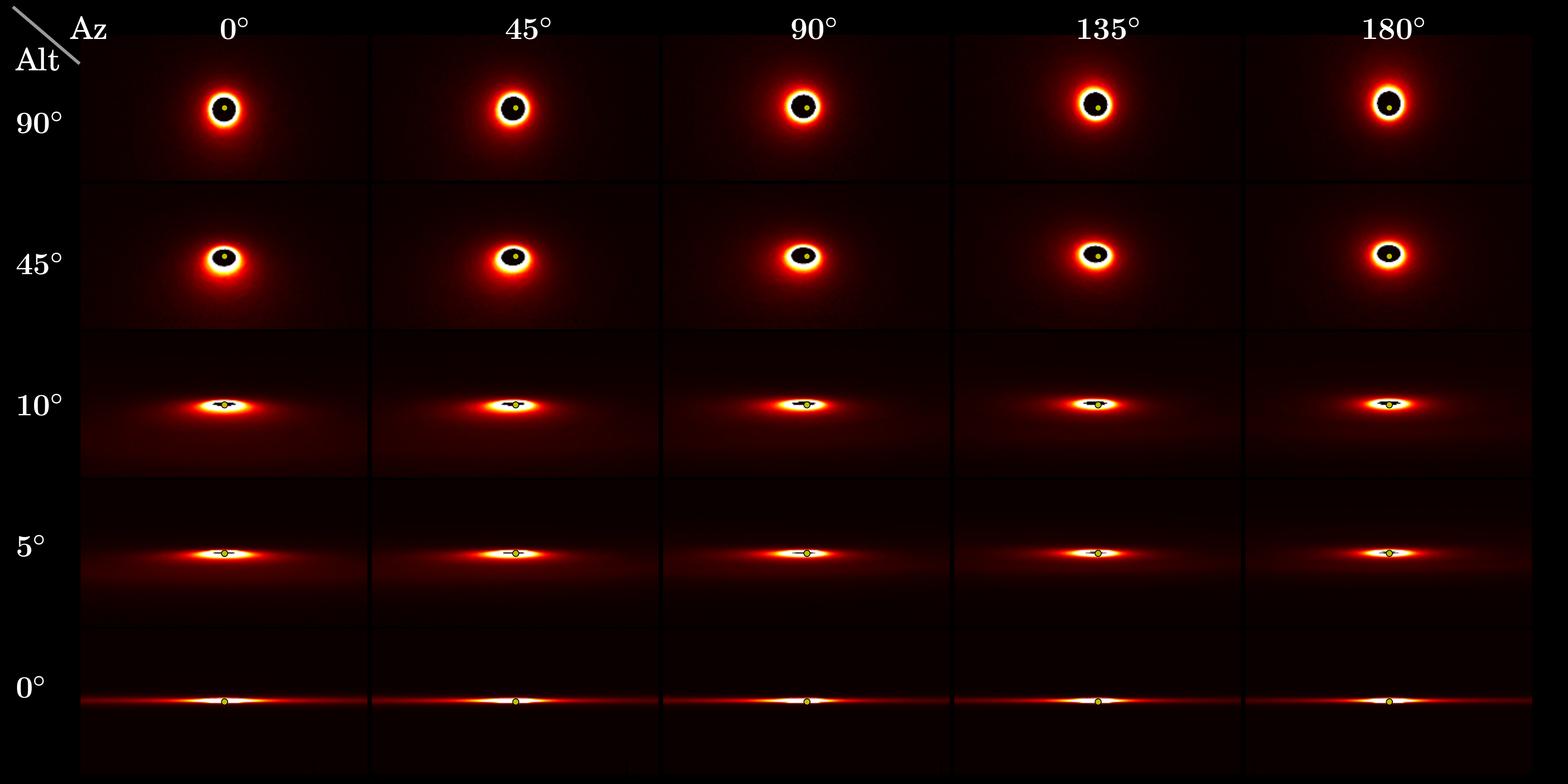}
    \caption{``Alt-az'' diagram for the case $e_{\rm planet} = 0.25$. 
    Synthetic scattered light images
    of the debris disk are shown
    as a function of the observer's altitude ($alt=0^\circ$/90$^\circ$ 
    gives an edge-on/pole-on view of the planet's orbit) and 
    azimuth ($az=0^\circ$/180$^\circ$ has the planet's apoapse/periapse 
    pointing toward the observer). For this and other alt-az figures, 
    we use an image scaling proportional to the 
    square root of the surface brightness.
    Each alt-az snapshot is constructed from an 800 AU $\times$ 800 AU
    grid, smoothed by convolving with a 2D Gaussian having a standard
    deviation of 2 AU, and truncated vertically to 400 AU. 
    The convolution shrinks
    the dust inner cavity; we restore
    the cavity seen in the pre-smoothed image 
    by masking out the corresponding pixels.
    The surface brightnesses of the brightest features are $\sim$600 ($10^4$) times 
    higher than that of the faintest features in the face-on (edge-on) view.
    The yellow dot in each panel marks the location of the central star.
    }
    \label{fig2}
\end{figure*}

\begin{figure*}[!ht]
    \includegraphics[width=\textwidth]{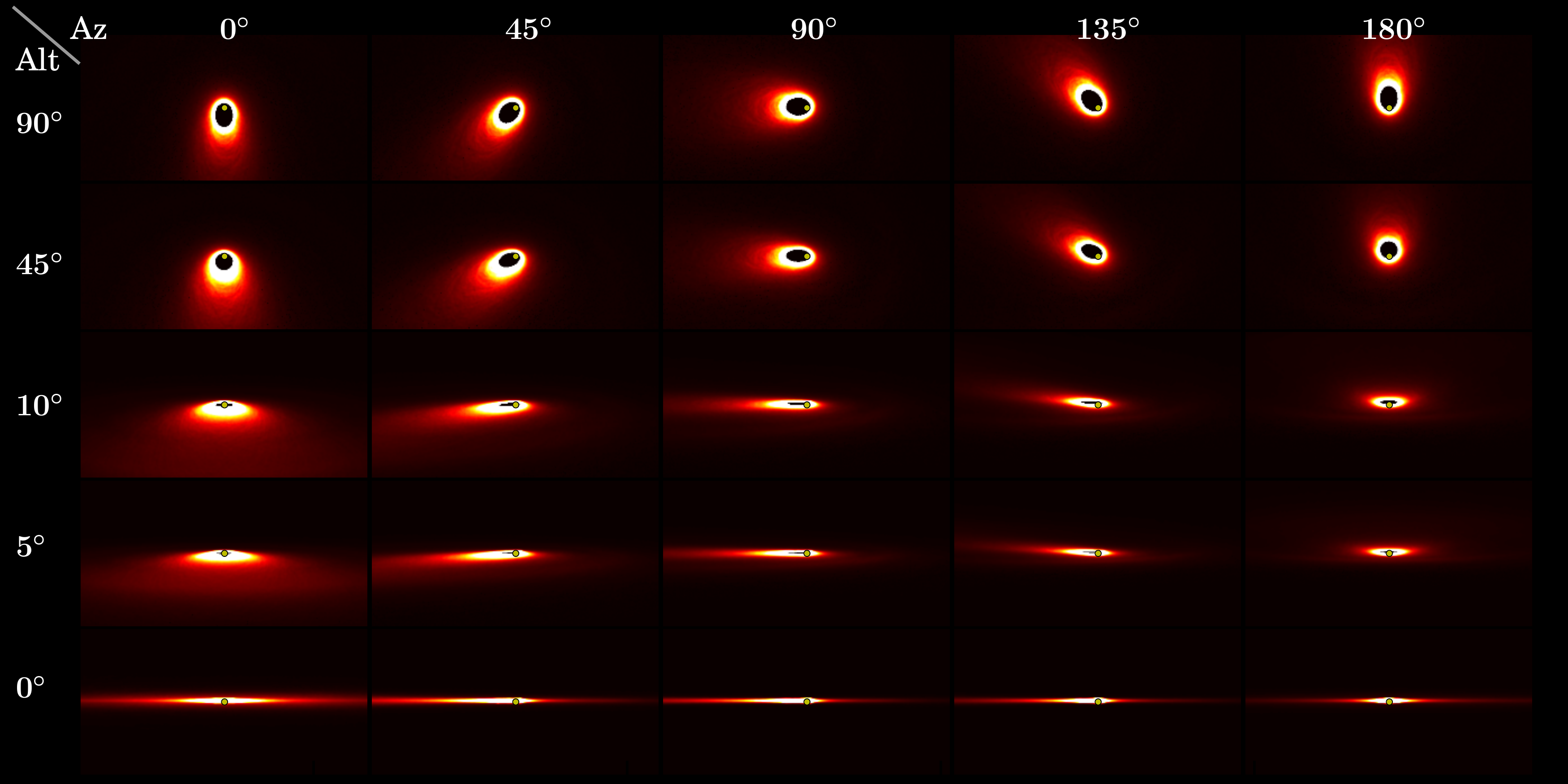}
    \caption{Same as Figure \ref{fig2}, but for a more 
    eccentric planet with $e_{\rm planet} = 0.7$. 
    The surface brightnesses of the brightest features are 
    $\sim$10$^4$ ($2\times 10^4$) times higher than that 
    of the faintest features in the face-on (edge-on) view.}
    \label{fig3}
\end{figure*}

\begin{figure*}
    \includegraphics[width=\textwidth]{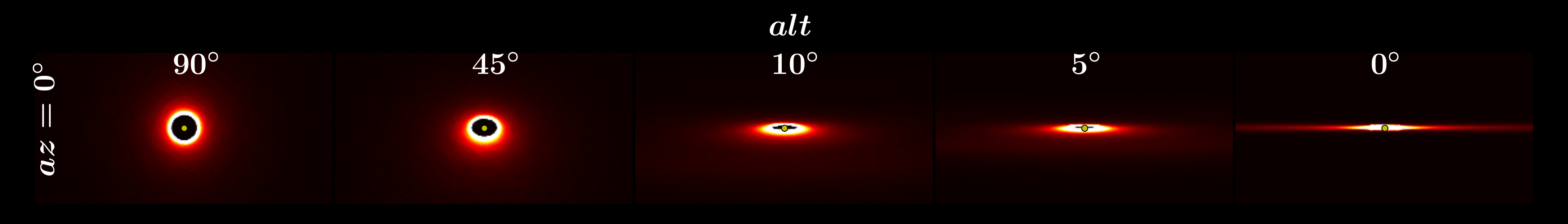}
    \caption{Same as Figure \ref{fig2}, but for $e_{\rm planet} = 0$.}
    \label{fig4}
\end{figure*}

\begin{figure*}
    \centering
    \includegraphics[width=\textwidth]{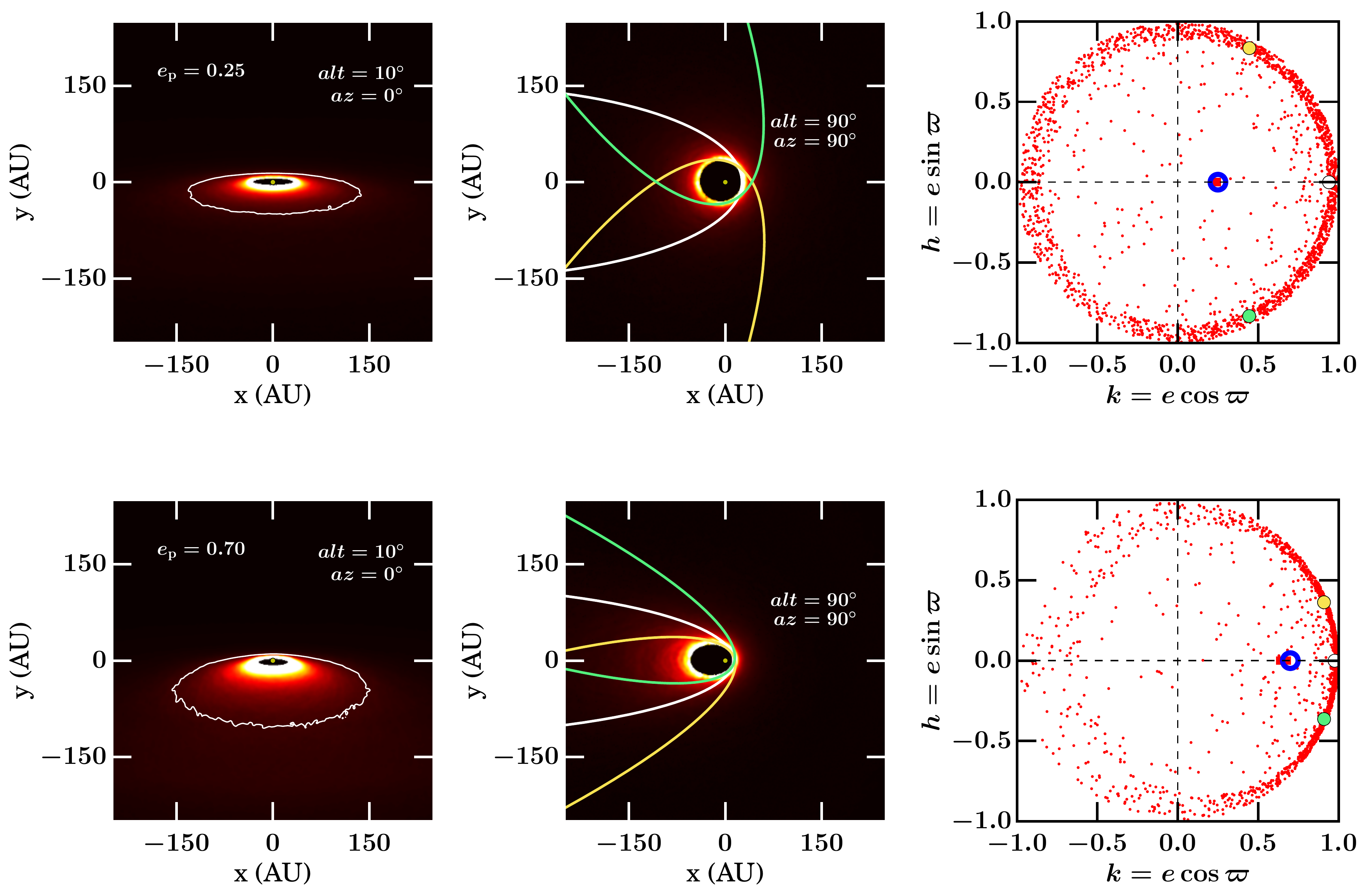}
    \caption{Zoomed-in images for $e_{\rm planet}=0.25$
    (top row)
    and $e_{\rm planet}=0.7$ (bottom row). Left panels are
    nearly edge-on views ($alt=10^\circ$), and are each 
    overlaid with a contour of constant surface brightness.
    Middle panels are face-on views showing representative
    dust grain orbits, 
    color-coded to correspond to the colored points in the
    right-hand $h$-$k$ plots (whose remaining symbols have
    meanings identical to those in Figure \ref{fig1}). 
    Note how most particles in $h$-$k$ space cluster near
    unit eccentricity as a consequence of our top-heavy
    $\beta$-distribution (\ref{eq:orbcorr}).
    The white dust orbit is launched from parent body
    periapse, and the green and yellow dust orbits are chosen to have
    median eccentricities and longitudes of periastron.
    As planet eccentricity increases, 
    increasingly many dust orbits have their periastra aligned
    with that of the planet, leading to a more extended
    and sharply angled ``fan'' of emission in nearly edge-on views.
   }
    \label{fig5}
\end{figure*}

Figure \ref{fig1} illustrates the basic ingredients of our model.
It depicts how the locations
of bodies in $(p,q,h,k)$ space relate to one another,
and to the resultant scattered light images,
for a sample case $e_{\rm planet} = 0.25$.
For pedagogic purposes, and for Figure \ref{fig1} only, 
we assign all dust particles a fixed $\beta = 0.4$,
discarding particles not bound to the star.
Surface brightness morphologies
can be understood in terms of underlying dust particle orbits
by starting from the 
face-on scattered light image (looking down the $z$-axis
onto the $x$-$y$ plane).
The inner dust cavity is outlined
by the launch sites of dust particles, i.e.,
the cavity rim coincides 
with the elliptical ring of parent bodies 
(the parent bodies themselves do not contribute 
to the scattered light image).
Because launch velocities are ``high'' for the weakened 
gravitational potential felt by dust particles (the potential is
weakened by $1-\beta$), the cavity rim / parent body ring
marks the periastra of dust particles. 
The bright half of the cavity rim,
located on the negative $x$-axis, traces the periastra
of $e \sim 0.3$ dust particles 
(drawn in white); these are launched from the apastra of their
parents' orbits. These same dust particles' apastra 
form the ``arc'' located to the right of the cavity.
The entire outer boundary
of dust emission, of which the arc is the brightest segment,
is demarcated by all the particle apastra.
Particles with the largest eccentricities
(e.g., yellow, blue, green, grey-brown), extending
up to unity, are launched
from near the periastra of their parents' orbits, and
form the barely visible ``wings''
extending above and below the arc. In a more edge-on
view, these wings increase in brightness because of their
increased line-of-sight optical depth.
Viewed at 5 deg above the planet's orbital plane, 
with the planet's apoapse directed toward the observer,
the wings appear downswept.

\section{Synthetic Scattered Light Images}
\label{sec3}

Figures \ref{fig2}, \ref{fig3}, and \ref{fig4} show the scattered
light images for $e_{\rm planet} = 0.25$,  
$e_{\rm planet} = 0.70$, and $e_{\rm planet} = 0$, respectively, with the 
radiation $\beta$ following a distribution given by (\ref{eq:orbcorr}). 
We smooth away some of the shot noise 
caused by a finite number of dust grains
by convolving images (from Figure \ref{fig2} onward)
with a 2D Gaussian having a standard deviation
of 2 pixels (2 AU). 
A side effect of the convolution is that 
it shrinks
the dust inner cavity; we restore the cavity
by masking out the corresponding pixels.
The panels in each figure are computed from a variety of vantage
points. The orientation of the observer (on the celestial
sphere centered on the debris disk) is parametrized
by altitude $alt$ (inclination angle relative to the planet's
orbital plane;
$alt = 0^\circ$ corresponds to the planet's orbit seen edge-on,
while $alt = 90^\circ$ gives a face-on view)
and azimuth $az$ (angle measured in the planet's orbital
plane; $az = 0^\circ$ corresponds to the apoapse of the planet's
orbit pointing toward the observer, while $az = 180^\circ$ directs
the planet's periapse toward the observer). 
For all images we rotate first in $alt$ and then in $az$
starting from $(alt,az) = (90^\circ, 0^\circ)$.
We refer to Figures \ref{fig2}--\ref{fig4} as ``alt-az'' diagrams.

All three alt-az diagrams are displayed on a universal
brightness scale.
To bring out the fainter features, images are scaled 
to the square root of the surface brightness. More edge-on views 
have greater line-of-sight optical depths and 
therefore yield brighter disks. 
For reference, the angular 
half-thickness of our disk is $\max i_{\rm p,free} = 0.02$ 
rad $\simeq$ 1$^\circ$.
Later in this section we will experiment with a thicker disk for which
$\max i_{\rm p,free} = 0.15$ rad $\simeq$ 9$^\circ$.

\begin{figure*}
    \centering
    \includegraphics[width=\textwidth]{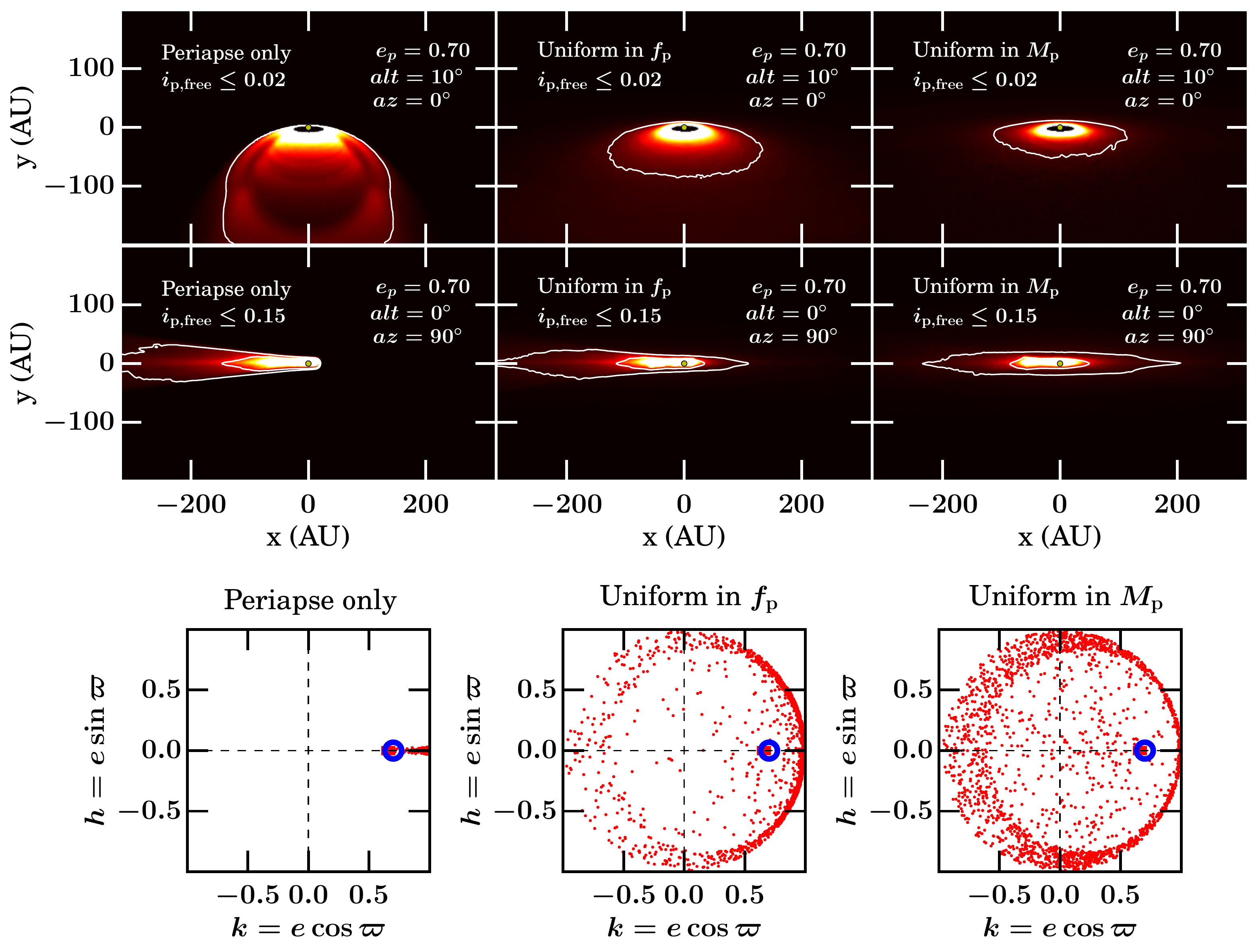}
    \caption{Experiments in the distribution of launch
    sites for dust particles, for the case 
    $e_{\rm planet} = 0.7$. If dust grains are launched
    strictly from the periastra of parent bodies, then all
    orbits are apsidally aligned 
    (left column of panels;
    symbol meanings in the $h$-$k$ plot
    are identical to those in Figure 
    \ref{fig1}). If dust grains are launched at parent body
    mean anomalies $M_{\rm p}$
    that are uniformly distributed between
    $0$ and $2\pi$, the preference for apsidal alignment is 
    muted (right column of panels). Our standard model assumes 
    that dust grains are launched at parent body
    true anomalies $f_{\rm p}$ that are
    uniformly distributed between $0$ and $2\pi$, and
    represents an intermediate case (middle column of panels).
    The top row displays corresponding scattered light
    images, observed at $alt = 10^\circ$
    and $az = 0^\circ$ for our standard vertically thin disk
    with $\max i_{\rm p,free} = 0.02$ rad. Each nearly edge-on disk, as
    traced by a white contour of constant surface brightness, 
    resembles a ``fan'' or ``moth''; the wings of the moth
    are angled more sharply downward as dust particle
    orbits are more strongly apsidally aligned (reading right to left).
    Note the ``double winged moth'' that appears when
    dust grains are launched exclusively from parent periastra (top left).
    The middle row features scattered light images observed
    at $alt = 0^\circ$ and $az = 90^\circ$ for
    a vertically thicker disk with $\max i_{\rm p,free} = 0.15$ rad.
    The center panel features an inner ``ship'' surrounded
    by its ``wake,'' as detailed in the main text.
    Brightness asymmetries and vertical asymmetries across the ship-and-wake are magnified as dust grain launch sites concentrate
    toward parent body periastra (reading right to left).}
    \label{fig6}
\end{figure*}

In many of the views displayed in Figures \ref{fig2}--\ref{fig3}, 
the eccentricity of the debris disk forced upon it by 
the eccentric planet manifests itself as
a stellocentric offset: the star is displaced from the apparent 
geometric center of the ring's inner cavity.
Another signature of planet eccentricity
is the tail of scattered light extending to one side of the star,
seen most prominently for high $e_{\rm planet}$. This tail 
arises from dust
particles on high-eccentricity orbits
launched from near the periastron
of the parent body ring;
in our diagnostic Figure \ref{fig5}, these orbits
are color-coded green, white, and yellow
(see in particular the bottom panels).
High-eccentricity
dust abounds as our $\beta$-distribution
(\ref{eq:orbcorr}) is strongly
weighted toward the maximum value just short of
radiation blow-out.
When this ``sheet'' of high-$e$ dust particles
is viewed just above its
plane ($alt \sim 5$--$10^\circ$) and in front of the star 
($az=0^\circ$),
it appears in projection as a ``fan'' or ``moth'' 
whose wings sweep downward from the
star. Higher planet eccentricities
cause both the top and bottom boundaries of the fan
to be more angled; compare the white
contours in the leftmost panels of Figure \ref{fig5}.
Maintaining the same above-the-plane view 
($alt \sim 5$--$10^\circ$),
but now rotating the observer in azimuth
so that the sheet of high-eccentricity dust
is seen behind the star ($az=180^\circ$),
produces an upswept fan (see Figures \ref{fig2}--\ref{fig3}).
Observer azimuths intermediate between $0^\circ$ and $180^\circ$ 
yield simultaneous top-down and left-right asymmetries.
For example, comparing the left and right limbs
of the disk seen at $az=135^\circ$ and $alt=10^\circ$ in 
Figure \ref{fig3}, we see that the left limb is more extended in
length, has a lower peak brightness, and is angled more upward.

When the planet's orbit is viewed nearly but not completely
edge-on ($0^\circ < alt \leq 10^\circ$), with its
periapse pointing toward the observer ($az > 90^\circ$),
a faint ``bar'' emerges displaced from the star. 
This bar, seen below the star in Figures \ref{fig2}--\ref{fig3},
is equivalent to the ``arc'' seen in front
of the planet's periastron in Figure \ref{fig1}; the bar/arc
is comprised of dust grains at their apastra,
on orbits launched from near the apastra of their parent
bodies. These orbits are apsidally anti-aligned
relative the planet's orbit (see the white orbit
in Figure \ref{fig1}). The bar is brightest
when seen in forward-scattered
light and at low observing altitudes which enhance its line-of-sight
optical depth.

The above mentioned tail of scattered light extends
only in the direction of the parent body disk's
apastron (and by extension the planet's apastron)
because there are more dust grains
on orbits apsidally aligned with their parents' orbits than anti-aligned.
This preference for apsidal alignment
magnifies with increasing planet eccentricity,
as shown in Figure \ref{fig5}, and can be understood
as follows.
For the simplifying case of coplanar orbits,
dust grains have 
$0 \leq |\varpi_{\rm d} - \varpi_{\rm p}| < \pi/2$ 
if they are launched 
between a parent body's periastron and its semi-minor vertex 
(where the semi-minor axis crosses the orbit). 
The range of 
true anomalies between periastron and 
the semi-minor vertex is 
greater than between the semi-minor vertex and apastron. 
This difference grows as $e_{\rm planet}$ grows; consequently,
more dust grain orbits have
$0 \leq |\varpi_{\rm d} - \varpi_{\rm p}| < \pi/2$ 
as $e_{\rm planet}$ increases.

The degree to which dust orbits
apsidally align with the parent body ring depends
not only on planet eccentricity, but also
the distribution of parent body true anomalies at launch.
Different distributions of launch sites are explored in Figure
\ref{fig6}. Alignment is perfect --- and the wings of the disk
seen in projection are swept most strongly downward --- if dust
grains are launched exclusively from periastron (left panels). 
If instead launch mean anomalies are uniformly distributed --- i.e., if 
launch true anomalies are weighted toward apastron where parent 
bodies linger --- then apsidal alignment is weakened (right panels).
Our standard model assumes a uniform distribution
of launch true anomalies and represents an intermediate case 
(middle panels).

\begin{figure}
    \centering
    \includegraphics[width=0.5\textwidth]{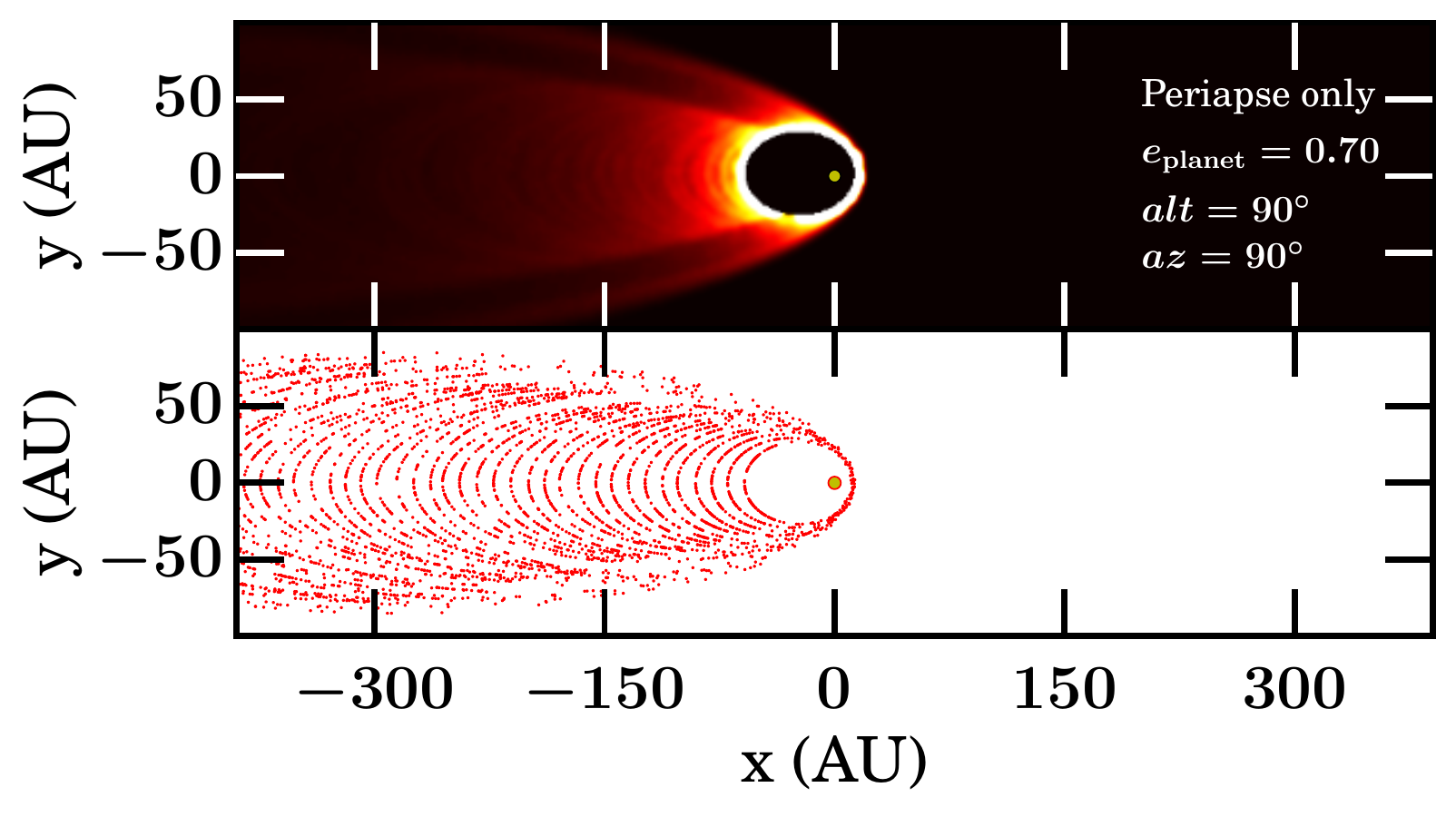}
    \caption{Understanding the origin of the double 
    wings seen for some moths, as seen in the upper left panel of Figure \ref{fig6} 
    (see also the middle right panel of Figure
    \ref{fig9} for another version of the double wing
    morphology).
    Double wings appear when all dust grains share practically the
    same periastron and apsidal line ($x$-axis)
    as a consequence of being launched
    only at parent body periastra.
    {\it Top:} Scattered light image of the same disk 
    featuring double 
    wings, but seen face on here.
    Emission near periastron generates the
    upper set of wings in Figure \ref{fig6}, while
    the pair of jet-like features
    displaced symmetrically above and below 
    the apsidal line produces the lower set of wings.
    {\it Bottom}: Same as top, but plotting individual
    dust grains. 
    Local overdensities generated at two orbital azimuths
    correspond to the two jets seen in the top panel.
    }
    \label{fig7}
\end{figure}

In the endmember case that all dust particles are launched
at parent body periastra and have their orbits completely
apsidally aligned, we can discern two sets of wings:
a thin pair of wings sitting above a more diffuse and roughly parallel
pair of wings below the star (top left panel of Figure \ref{fig6}).
We can understand this ``double wing'' morphology
using the face-on views shown in Figure \ref{fig7}.
The upper set of wings seen in Figure \ref{fig6}
corresponds to the bright arc near periastron in Figure \ref{fig7}.
This arc is especially luminous because of the confluence
of orbits converging on nearly the same periastron.
The lower set of wings in Figure \ref{fig6}
corresponds in Figure \ref{fig7}
to the pair of overdense ``rays'' located
toward parent body apastron and
symmetrically displaced above and below 
the apsidal line (the $x$-axis).
These two local maxima in surface brightness --- 
what look like a pair of jets spraying particles away from the star
in the face-on view --- arise from two
effects: (1) the tendency of particles
on a given orbit to be found closer
to apoapse (where they linger) than to periapse
(which they zip through), and (2) 
the lowering of the particle density 
along the apsidal line in the direction of apastron,
due to large orbit-to-orbit differences in
apastron distance --- see the
bottom panel of Figure \ref{fig7}. The broad distribution
of apastron distances (extending to infinity) is due in turn to
a radiation-$\beta$ distribution that abuts blow out.
Effect (1) concentrates particles 
toward apoapse,
while effect (2) dilutes the particle density 
along the apsidal line (in the direction of apastron); 
the net effect is to concentrate particles
at two orbital phases symmetrically displaced away from the
apsidal line.

The difference between the ``double wing'' and the ``bar'' 
lies in their relative proximities to the central star.
The outermost wing --- what defines the edge of the disk --- cuts
almost directly across the star in projection, whereas the
bar is necessarily displaced from the star.

\begin{figure}
    \centering
    \includegraphics[width=0.5\textwidth]{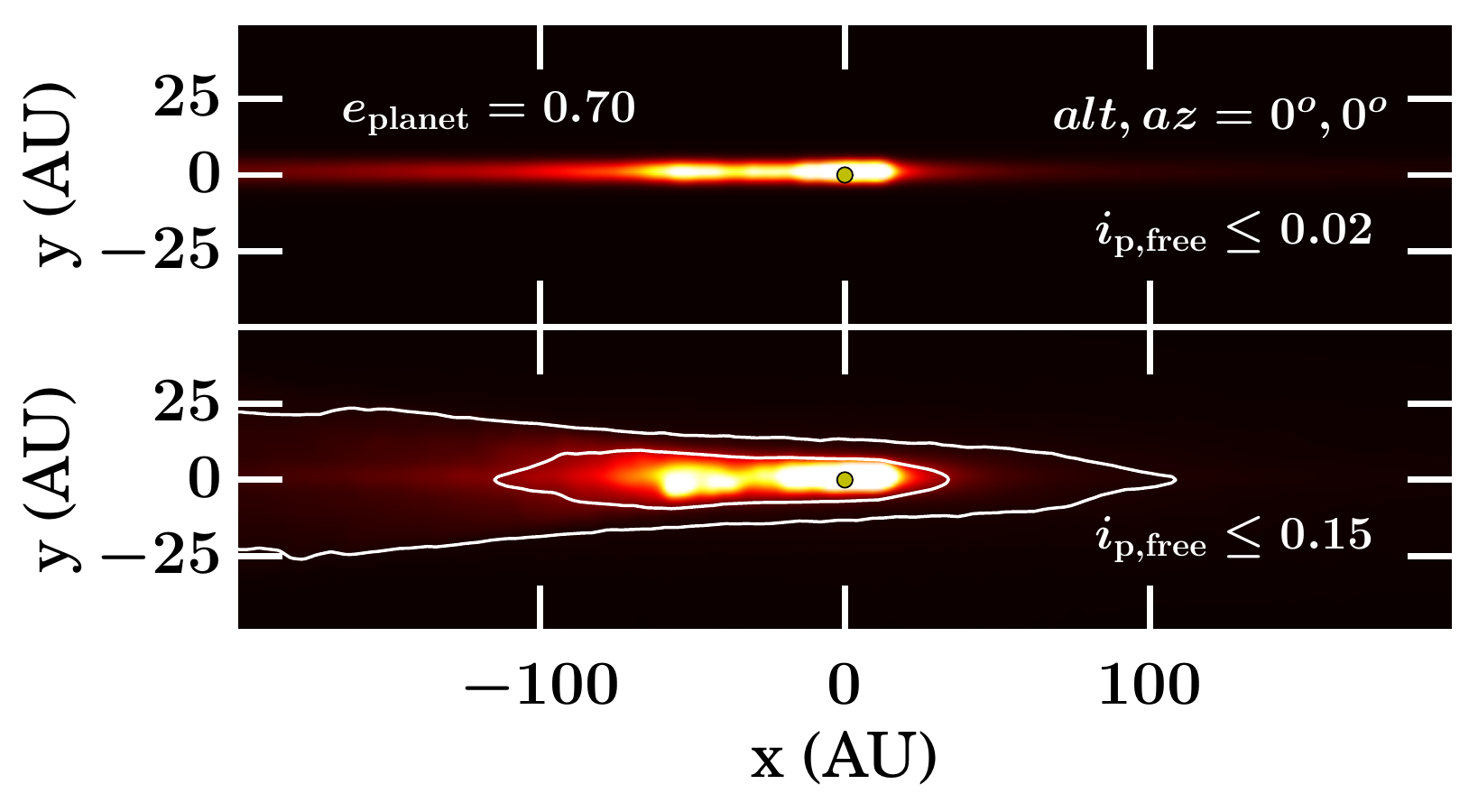}
    \caption{A sufficiently thick disk ($\max i_{\rm p,free} = 0.15$ rad;
    bottom panel) seen edge-on
    features a ``ship'' (inner white contour of constant
    surface brightness) and its surrounding ``wake''
    (outer white contour). The ship's front/bow (on
    the positive $x$-axis, aligned with the underlying planet's
    periastron) is brighter than its back/stern. 
    The outer wake is narrower at its front than its back.
    The ship-and-wake morphology
    might be relevant for HD 106906; see Section \ref{sec4}.}
    \label{fig8}
\end{figure}

All of the behavior reported above persists if $\max i_{\rm p,free}$
is increased from our standard value of 0.02 rad to 0.15 rad;
i.e., the alt-az diagram for a vertically thicker disk looks similar
to that of our standard thin disk.
But there is more. 
Thickening the disk, and viewing it
edge-on ($alt=0^\circ$) 
and near quadrature ($45^\circ \leq az \leq 135^\circ$),
reveals new morphological features, as seen
in Figure \ref{fig8}. 
The disk's outer isophote 
has a front (toward planet periastron)
that is vertically thinner than its back, 
resembling the ``wake'' of a ``ship'' (inner isophote
enclosing the dust cavity rim seen in projection).
The head of the ship
and the back of its wake
comprise dust on orbits that are highly eccentric
and closely apsidally aligned with the parent disk
(these are represented by the white, green, and yellow orbits
in Figure \ref{fig5}).
Conversely, the 
stern of the ship
and the front of its wake
coincide with the few 
dust orbits that are 
anti-aligned with the parent disk 
and less eccentric.
The wake grows in vertical thickness from front to back
because the front is composed of dust at the apastra of
low eccentricity orbits,
while the back is composed of dust at the more distant apastra
of high eccentricity orbits;
at fixed inclination dispersion, the more
distant apastra have greater heights above the disk midplane.
As was the case for the moth (see above),
the degree of 
vertical asymmetry for the 
wake depends on the distribution of dust grain launch sites: 
the more the launch sites concentrate near periastra, 
the more severe the asymmetry (see middle row of
Figure \ref{fig6}).

\begin{figure*}
    \centering
    \includegraphics[width=\textwidth]{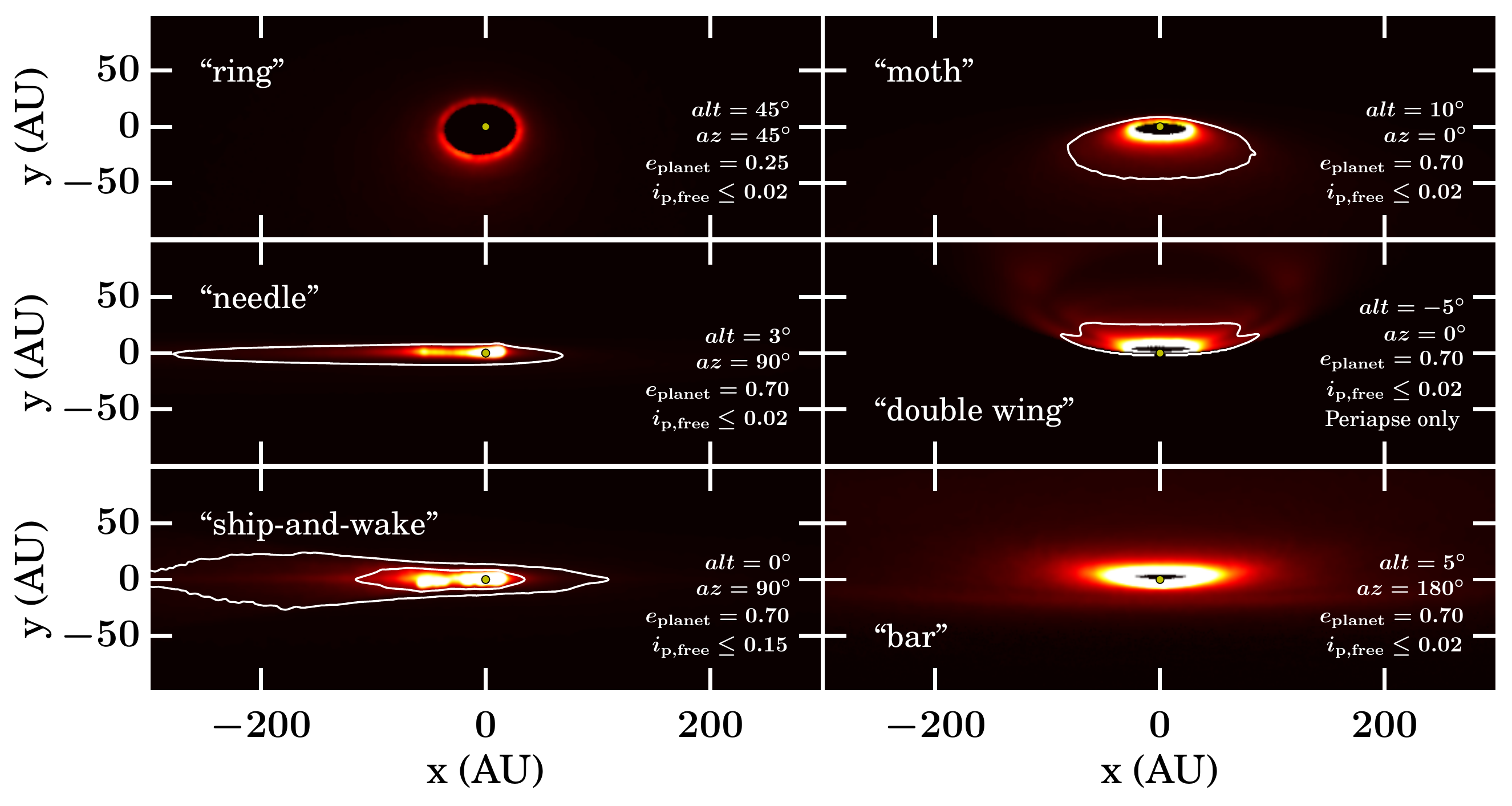}
    \caption{Prototypical debris disk morphologies
    seen in scattered light, as captured by a ``minimum model''
    (single eccentric planet + ring
    of parent bodies + dust grains + stellar radiation pressure). 
    Possible observable shapes include a ``ring'' (top left),
    a ``needle'' (middle left; this is essentially a ring seen edge-on),
    and a ``ship-and-wake''
    (bottom left; this is basically
    a needle which is fat enough to resolve vertical structure).
    Right panels feature various kinds of ``moths,'' either
    our standard version where most dust grains are in front
    of the star and therefore appear bright in forward-scattered
    light (top right), a moth with ``double wings'' where dust grain
    orbits are perfectly apsidally aligned as a consequence of assuming
    that grains are launched exclusively from parent body periastra
    (middle right), and a ``reverse moth'' where most grains are behind
    the star, accompanied by a ``bar'' in front of the star (bottom right).
    Note the sharp wingtips seen in the ``double wing'' panel;
    this model looks encouragingly similar to HD 32297
    \citep[][their Figure 19b]{schneider14}.
    The surface brightness contrasts between the brightest 
    and the faintest features are $\sim$36, $\sim$900, 
    $\sim$10$^4$, $\sim$260, $\sim$620, and $\sim$400 for 
    the ring, needle, ship-and-wake, moth, double wing, and 
    bar, respectively.
    The head of the ship is $\sim$400$\times$ brighter 
    than its stern.
    In the double wing, the two wings are $\sim$4$\times$
    brighter than the gap between them. 
    The bar is $\sim$20\% brighter than the gap 
    that separates it from the main disk.
    }
    \label{fig9}
\end{figure*}

\section{Summary and Discussion}
\label{sec4}

We have explored in this work what a ``minimum model'' of a
debris disk looks like in scattered light.
The minimum model consists of a narrow ring of parent bodies,
secularly perturbed by a single, possibly eccentric
planet, producing dust grains whose orbits are made arbitrarily
eccentric by stellar radiation pressure. The model
has obvious relevance to systems like Fomalhaut and HR 4796A
which patently feature narrow elliptical 
rings.\footnote{\citet{perrin15}
suggest that the HR 4796A disk
may be slightly optically thick.}
What might not be
so obvious is that the minimum model can also help to explain
many other morphologies documented in resolved images of debris disks ---
all by simple changes in viewing perspective.
A message emerging from our work is that the 
outskirts of planetary systems
are shaped by eccentric planets, 
possibly just a few Earth masses each.

In Figure \ref{fig9} we summarize the various disk shapes
that are possible. We classify these into five types:
``ring,'' ``moth,'' ``bar,'' ``needle,'' and ``ship-and-wake.''
The first four shapes can be generated even by a disk
that is completely flat. We review each of these morphologies
in turn, highlighting potential applications to observed systems,
and close by listing future modeling directions.

\subsection{``Ring''}
Dust that is generated from an eccentric ring of parent bodies
appears as an eccentric ring itself 
when viewed close to face on (top left panel of Figure \ref{fig9}).
The inner rim of the ring is illuminated by
dust particles near their periastra, while a skirt of diffuse
emission extends outward from dust grains
en route to their apastra.
Some real-life examples of rings with offset host stars
are provided by Fomalhaut
\citep[e.g.,][their Figure 1]{kalas05fom}, 
HR 4796A (e.g., \citealt{schneider09}, their Figure 3;
\citealt{thalmann11}, their Figure 1; 
\citealt{perrin15}, their Figure 8;
\citealt{grady16}),
HD 181327 \citep[e.g.,][their Figure 33]{schneider14},
and HD 157587 (\citealt{padgett16}; Millar-Blanchaer et al., submitted).
These systems also feature diffuse emission exterior to 
their rings.

\subsection{``Moth''}
When the parent body ring is viewed nearly but not completely
edge-on, with its apoapse pointing out of the sky plane
toward the observer,
a shape like a fan or moth materializes
(top right panel of Figure \ref{fig9}).
The resemblance of this morphology to the actual ``Moth''
(HD 61005; \citealt{hines07})
was first pointed out by \citet{fitz11}
and explored with detailed and quantitative models fitted
to the Moth by Esposito et al.~(submitted).
For sample images of HD 61005, see, e.g., Figure 3 of \citet{hines07};
Figure 1 of \citet{maness09};
Figure 1 of \citet{buenzli10};
and Figure 1 of \citet{ricarte13}.
The wings of our model moth are composed of
dust grains on highly eccentric orbits that are apsidally aligned with
the parent ring (and by extension the planet),
and whose apastra are directed toward the observer.
Viewing these grains from slightly above their orbital plane
produces downswept wings; viewing them from below produces upswept wings
(flip the top right panel of Figure \ref{fig9} about the $y$-axis).
If instead these grains' apastra are directed into the sky plane
away from the observer,
then the wings of the moth appear foreshortened because
most of the starlight is forward-scattered away from the observer 
(this is the ``reverse moth'' featured in the 
bottom right panel of Figure \ref{fig9}; the
foreshortening is not apparent because the panel
is made using a low contrast to highlight
another feature, the ``bar,'' which will be
discussed below).
Note that the moth morphology does not depend on a non-zero
inclination between the parent body ring and the planet;
a perfectly flat system suffices, provided it is viewed
slightly away from edge-on.

The degree to which the wings of the moth are angled
depends on the degree to which dust grain orbits are
apsidally aligned.
In turn, the preference for apsidal alignment depends 
on both planet eccentricity and the orbital phases
at which parents give birth to dust grains.
If dust grains are launched
from parent body periastra and no other orbital phase,
then the system is, in a sense, maximally non-axisymmetric;
there is a ``preferred'' direction in space;
apsidal alignment is perfect, and the moth wings
sweep most strongly away from the horizontal.
The wings of HD 61005 are angled
by $\sim$23 degrees from the horizontal
(\citealt{buenzli10}; Esposito et al., submitted),
suggesting high planet eccentricity and
a strong preference for 
launching dust grains near parent periastra.

Another moth-like system is presented by
HD 32297. Intriguingly, HD 32297 sports
a second, fainter pair of moth wings 
that roughly parallel the first,
as imaged by {\it HST} on scales of several arcseconds 
(e.g., \citealt{schneider14}, their Figures 18 and 19).
Our minimum model can reproduce this ``double wing''
structure (middle right panel of Figure \ref{fig9}).
When dust orbits are closely apsidally aligned,
a first set of wings (closest to the star, toward the bottom
of the panel) traces particles at and around their
periastra, while another fainter set of wings (farther from
the star, toward the top of the panel)
is generated by particles near, but not at, their apastra.
We can even try to make a connection to the disk geometry
as revealed on smaller, subarcsecond scales at infrared 
wavelengths. Figure 4 of \citet{esposito14}
(see also Figure 1 of \citealt{currie12})
reveals the emission closest to the star 
to be concave down (when north-west is up)
and the emission farther from the star to be concave up
(the latter curvature is consistent with the {\it HST} images 
from \citealt{schneider14}).
We can reproduce this reversal of concavity between
small and large scales 
by identifying the observed concave downward disk with
the bright arc above the star (the apoapse
of the innermost cavity rim, pointed toward
the observer), 
and the concave upward disk with 
the wingtips.

A third example of a fan/moth
is given by HD 15745; see, e.g., Figure 1 of
\citet{kalas07_hd15745} 
and Figures 13 and 14 of Schneider et
al.~(\citeyear{schneider14}; note the typo
in the source HD number in the caption to Figure 13).
Unlike the case for HD 61005 and HD 32297,
isophotal ellipses describe well 
the fan of HD 15745, and indicate that this disk is not
necessarily eccentric: an axisymmetric disk viewed somewhat
above its orbital plane, composed of grains that strongly
forward-scatter, can
reproduce the morphology of HD 15745. See Figure 4 of
\citet{kalas07_hd15745}, or our Figure \ref{fig4} (e.g., $alt = 10^\circ$).

\begin{figure}
    \centering
    \includegraphics[width=0.5\textwidth]{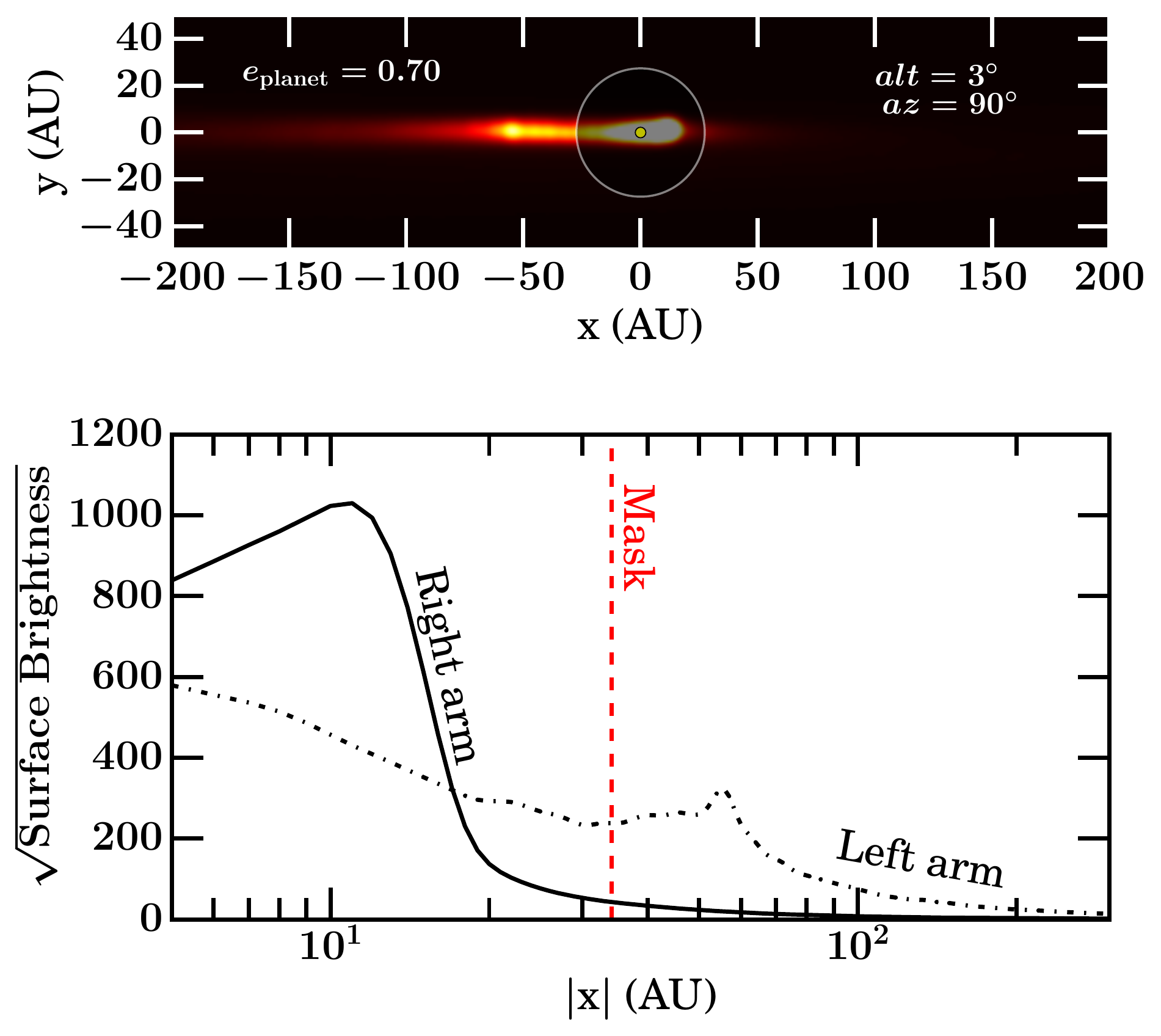}
    \caption{Zoom-in on our model ``needle.'' As long as the coronagraphic
    mask covers enough of the central cavity --- specifically the region near
    periapse, where the disk is at maximum brightness --- then the disk's longer
    arm can appear brighter than its shorter arm, as is consistent with
    observations of HD 15115. Accompanying surface brightness profiles for each
    arm are computed versus radius $|x|$ by integrating over $y$. Each profile features a local maximum where the line of sight intersects
    regions near the ansa of the cavity rim.
    }
    \label{fig10}
\end{figure}

An alternative way to produce a moth-like morphology is to
allow the interstellar medium (ISM) to secularly perturb dust grain orbits
\citep{maness09}.\footnote{Secular ISM perturbations on grains that
remain bound to the host star, as proposed by \citet{maness09},
should not be confused with ISM deflections of
unbound grains \citep{debes09}.
Unbound grains contribute negligibly to disk
surface brightness; compared to bound grains,
unbound grains have lifetimes that are shorter by orders of magnitude,
and so their relative steady-state population
is correspondingly smaller
\citep[e.g.,][]{strubbe06,krivov06}.
See \citet[][their section 4]{maness09} for a detailed
discussion of the various flavors of ISM interactions,
including empirical arguments against interaction with
a high density ($\sim$100 atoms per cm$^3$)
ISM in the case of HD 61005.}
The mono-directional flow of the ISM across the disk
can induce a global disk eccentricity and thereby mimic
some of the effects of an eccentric planet.
As \citet{maness09} recognize (see their section 5.1),
this mechanism is subject to uncertainties in the host stellar wind;
in principle, the stellar wind can blow
an ``astrosphere'' shielding disk grains from ISM interactions.

\subsection{``Bar''}
A faint bar emerges when disks are viewed close to
but not completely edge-on, 
with the embedded planet's periapse pointing out 
of the sky plane
(bottom right panel of Figure \ref{fig9}).
The bar, which can be $\sim$20\% brighter than the
gap separating it from the main disk,
is composed of dust grains lingering at the apastra
of orbits that are nearly apsidally anti-aligned
relative to the planet's orbit. 
These grains are launched
onto highly eccentric, barely bound orbits
from the apastra of the parent body ring. 
Detecting the bar
would confirm that the grain size distribution rises sharply
toward the radiation blow-out value as a consequence 
of the long collisional lifetimes afforded by highly eccentric
grains \citep{strubbe06}. 
Such a top-heavy size distribution ensures that
dust orbit eccentricities cluster about a unique value;
a pure Dohnanyi size distribution is actually insufficiently
top heavy and does not produce bars.

\subsection{``Needle''}
Needles appear when eccentric and vertically thin disks are viewed
edge-on with their semi-minor axes nearly parallel to the line of sight
(middle left panel of Figure \ref{fig9}).
Needles possess not only length asymmetries --- one limb
appears longer than the other --- but also brightness
asymmetries.\footnote{Of course, if the parent
ring is circular, or if an eccentric ring is seen exactly edge-on with
its major axis parallel to the line of sight, then
both limbs will appear of equal length and brightness (Figure \ref{fig4},
$alt=0^\circ$).
This limiting case of a ``symmetric needle'' may apply to
AU Mic, modulo its mysterious
non-axisymmetric and time-dependent clumps 
\citep{fitzgerald07,schneider14,wang15,boccaletti15}.}
As Figure \ref{fig10} details, the shorter arm, containing
dust grains crammed closer to the star, has a higher peak brightness
where the line of sight runs through the periapse of the ring cavity.
Our model needle is reminiscent of the prototype
HD 15115 (``The Blue Needle''):
see, e.g., Figure 1 of \citet{kalas07}; Figure 11 of 
\citet{schneider14}; 
and Figure 1 of \citet{macgregor15}.
These observations show the longer arm to be brighter than the shorter
arm (cf.~Figure 1 of \citealt{rodigas12} and Figure 1 of 
\citealt{mazoyer14}
which show more of a brightness asymmetry than a length asymmetry).
Bright long arms can be explained by our model needle provided
the coronagraphic mask is large enough to block out
the global maximum in surface brightness which lies
along the shorter arm; see Figure \ref{fig10}.
A prediction of the model is that beneath the mask,
the surface brightness profiles of the two arms
should cross, with the shorter arm ultimately outshining
the longer arm sufficiently close to the star.

\subsection{``Ship-and-Wake''}
Akin to needles are ships and their associated wakes,
which appear when
eccentric parent rings, viewed edge-on 
and close to quadrature,
have sufficiently large inclination
dispersions that vertical structure can be resolved
(bottom left panel of Figure \ref{fig9}).
The ship appears on length scales of the inner cavity rim.
The wake, tracing large-scale diffuse emission,
is vertically thicker 
in the direction of the planet's apastron.

The wake might be relevant for HD 106906.
On comparatively small scales within $\sim$1 arcsec (92 AU) of the star,
the disk's western arm appears shorter 
than its eastern
arm, as resolved by the {\it Gemini Planet Imager} and {\it SPHERE}
(Figure 1 of \citealt{kalas15}
and Figure 1 of \citealt{lagrange16}, respectively). 
We would interpret these observations
to imply that the underlying planet's periapse points west.
On larger scales outside $\sim$2 arcsec, the {\it Hubble Space
Telescope (HST)} reveals the nebulosity to the
east to be more diffuse than to the west (\citealt{kalas15},
their Figure 3) --- this is consistent with the eastern nebulosity
being the back of the wake, comprising dust grains
near the apastra of eccentric orbits apsidally aligned
with the planet's.
A potential problem with this interpretation
is that the {\it HST} image also evinces a radially long
extension to the west, suggesting that apoapse points
west instead of east.
The complete picture
must ultimately include HD 106906b, the substellar
companion at a projected distance of $\sim$7 arcsec 
from the star \citep{bailey14}.
It may be that the system is not dynamically relaxed
but has been perturbed by a flyby \citep{larwood01,kalas15}.

\subsection{Future Improvements}\label{fi}
Our model can be improved in a number of ways.
A more accurate calculation of
the distribution of dust grain launch sites as a function
of parent body orbital phase would be welcome.
We found that the appearance of ``moth''-like disks depended
on this distribution: if parent bodies collide preferentially
near their periastra, launching more dust grains there,
then the wings of the moth would be angled more sharply downward.
Collision rates and grain size distributions, each a function
of position, depend on one another; moreover, the entire
disk is spatially interconnected, as dust grains on orbits made
highly eccentric by radiation pressure can collide with
particles at a range of orbital radii.
Numerical simulations --- e.g., {\tt SMACK}
\citep{nesvold13}, augmented to include radiation
pressure \citep{nesvold15} --- can help to solve this problem.

The impact of different scattering phase functions
can be explored. Our images, constructed with a 
single Henyey-Greenstein scattering phase function
having a fixed asymmetry parameter $g$,
can be made more realistic
by accounting for how smaller grains
scatter light more isotropically (smaller grains should
have smaller $g$ values than larger grains).
\citet{hedman15} find empirically that
the light scattering properties of Saturn's rings
resemble those of irregularly shaped particles
and submit them for application to debris disks.

Warps --- misalignments between inner and outer disks --- 
are missing from our models of single planets in steady-state 
(secularly relaxed) disks. Positing two or more planets
on mutually inclined orbits produces warps 
\citep[e.g.,][]{wyatt99}. A single planet can also 
induce a transient warp \citep{mouillet97},
as has been apparently confirmed by the discovery of
beta Pictoris b (\citealt{lagrange10}). See also, however,
\citet{mmb15} and \citet{apai15}
who report features in beta Pic that a single planet may be
unable to explain.

Higher-order secular effects relevant at high planet eccentricity
and high inclination relative to the parent body disk
\citep[e.g.,][]{veras07,li14,pearce14,nesvold16},
and explicit numerical tests of dynamical stability, 
can also be incorporated in future models.
Other neglected dynamical effects include those
from non-overlapping mean-motion resonances
\citep[e.g.,][]{kuchner03,stark09}.
Observational evidence for the relevance of individual
MMRs is so far scant except for
among Kuiper belt objects \citep[e.g.,][]{batygin16,volk16}
and in the inner Solar System's zodiacal dust disk 
\citep[e.g.,][]{dermott94,reach10,jones13}.

\acknowledgments
We thank Gaspard Duch\^ene, Tom Esposito, Mike Fitzgerald, 
James Graham, Paul Kalas, Max Millar-Blanchaer,
Ruth Murray-Clay, Erika Nesvold, Chris Stark,
and Jason Wang for
encouraging and insightful discussions, and prompt and constructive
feedback on a draft version of this manuscript.
An anonymous referee provided a helpful and exceptionally fast report.
EJL is supported in part by the Natural Sciences and 
Engineering Research Council of Canada under PGS D3 and 
the Berkeley Fellowship. EC acknowledges support from 
grants AST-0909210 and AST-1411954 awarded by the 
National Science Foundation, and NASA Origins grant 
NNX13AI57G.
This research used the Savio computational cluster 
resource provided by the Berkeley Research Computing 
program at the University of California, 
Berkeley (supported by the UC Berkeley Chancellor, 
Vice Chancellor of Research, and the
Chief Information Officer).

\bibliography{debris}
\end{document}